\documentclass[conference]{IEEEtran}
\IEEEoverridecommandlockouts
\usepackage{cite}
\usepackage{amsmath,amssymb,amsfonts}
\usepackage{algorithmic}
\usepackage{booktabs}
\usepackage{graphicx}
\usepackage{textcomp}
\usepackage{multirow}
\usepackage[table]{xcolor}
\def\BibTeX{{\rm B\kern-.05em{\sc i\kern-.025em b}\kern-.08em
    T\kern-.1667em\lower.7ex\hbox{E}\kern-.125emX}}
\begin{document}

\title{
Understanding Everyday Communication Patterns, Stress, and Productivity in Hybrid Work through Egocentric Wearable Audio Sensing
}

\author{
\IEEEauthorblockN{
Tiantian Feng$^{\star}$, Aditya Kommineni$^{\star}$, Emily Zhou$^{\star}$, Burcin Becerik-Gerber$^{\ddagger}$,\\
Gale Lucas$^{\dagger}$, Shawn C. Roll$^{\dagger}$, and Shrikanth Narayanan$^{\star}$
}
\\
\IEEEauthorblockA{Signal Analysis and Interpretation Laboratory, University of Southern California, USA$^{\star}$\\ 
Chan Division of Occupational Science and Occupational Therapy, University of Southern California, USA$^{\dagger}$ \\
Sonny Astani Department of Civil and Environmental Engineering, University of Southern California, USA$^{\ddagger}$}
\thanks{This work was supported by NSF under Grant No. IIS-2204942.}
}

\maketitle

\begin{abstract}
The hybrid work trend in the post-COVID period has transformed how people communicate while at work. 
However, there is a lack of quantitative understanding of how home and office environments shape communicative behavior and affective well-being.
To study this, as a part of a larger study of the role of work environment on stress, we designed an egocentric wearable audio sensing system that captured interpersonal communication across both working locations. 
The system extracts real-time acoustic features on a wearable audio sensor and performs offline foreground speaker diarization to derive speaking load and activeness features. 
To improve diarization performance under low-dimensional feature constraints, we apply a teacher-student knowledge distillation framework that transfers knowledge from large-scale speech foundation models into a compact ResNet-based model, achieving consistent improvements in diarization performance over the baseline across three public egocentric speech datasets.
In a four-month longitudinal deployment with 16 participants, our system revealed location-dependent differences in work communication patterns. Moreover, the mixed linear model analyses showed that office-based speaking load was significantly associated with higher self-reported stress and lower self-reported productivity, while speaking activeness during home-based work amplified both productivity and affective load in this cohort.
\end{abstract}


\section{Introduction}

The COVID-19 pandemic has substantially reshaped where and how people work, leading to a major shift from traditional office-based work to fully remote or hybrid working environments. These changes have introduced meaningful differences in everyday work behaviors across working locations, including variations in physical activity, communication patterns, and overall work productivity~\cite{fukushima2021associations, koohsari2021changes, awada2021working}. Office-based work is typically situated within a structured physical and social environment, where movement between spaces, scheduled meetings, and communal meal times help segment the workday. On the other hand, those working from home often encounter more prolonged sedentary periods, household interruptions, and a blurring of boundaries between work and personal life~\cite{fukumura2021negotiating}. Therefore, understanding how these behavioral differences persist across home and office settings is critical for characterizing how work practices and daily routines have evolved since the wide adoption of remote or hybrid working.

In particular, apart from physical and productivity-related differences, two distinct environments are naturally associated with unique patterns of communication and social engagement. In the office, typical social communication tends to be spontaneous or scheduled interactions with colleagues, conversations with customers, and shared organizational gatherings. In contrast, working from home typically involves more scheduled virtual meetings and decreased informal exchange~\cite{viererbl2022you}. By comparing communication behaviors across these two working environments, this study aims to provide empirical insight into how hybrid work arrangements in the post-COVID period are related to everyday communicative activity at work.

Two fundamental features that characterize social communication are the frequency and duration of conversational interactions. For example, prior research has shown that frequent communication is critical for developing group cohesiveness, coordinating work activities, and increasing interpersonal trust within teams~\cite{walther2005rules}. Specifically, the work location plays a critical role in shaping both dimensions. In office environments, physical co-location naturally offers rich communicative opportunities, from spontaneous hallway exchanges to collaborative sessions in shared spaces. In contrast, remote work tends to transition synchronous communication into asynchronous interactions, resulting in more static and siloed collaboration networks~\cite{yang2022effects}.
Therefore, these differences in communication at work may carry implications for individual productivity, stress, and well-being. To understand such differences through continuous and in-the-wild measurement, this work designs and applies an egocentric wearable audio sensing system to unobtrusively capture everyday communicative activity across home and office environments while at work. 

Specifically, sensing everyday communication patterns poses two key challenges.
The first challenge is to ensure device comfort, scalability, and privacy-enhanced design. To address this, we build on our previously deployed wearable audio sensor, called TAR~\cite{feng2018tiles}, which integrates a lightweight, budget-friendly smartphone as the sensing device. This device has been deployed in a 10-week study to understand the communication patterns of more than 300 hospital caregivers~\cite{yau2022tiles, mundnich2020tiles2018}. 
A customized application performs real-time on-device feature extraction, capturing only a limited set of prosodic features and MFCCs (Mel-Frequency Cepstral Coefficients) to reduce privacy risks.
The second challenge is identifying speech regions belonging to the device wearer to derive reliable communication features with limited low-level features. We address this with a foreground speaker detection model that distills knowledge from speech foundation models to recognize the wearer's voice using only a small set of MFCCs and improve noise robustness by augmenting training data with indoor environmental noise.

The contributions of this work are as follows: 1) We deployed an end-to-end egocentric wearable audio sensing system in a higher education institution to study everyday communication patterns among office workers, including administrators, research faculty, and student service managers. The wearable audio sensor performed on-device processing to transform raw audio signals into low-level acoustic features. 2) To filter the communication activity of the device-wearing participant, we developed an effective speaker detection model that relied on a limited MFCC set and identified speaking regions corresponding to the device wearer. Based on the retained acoustic features from the wearer, we derived interpretable behavioral measures, such as speaking load and activeness, to characterize everyday communication patterns.
3) Based on a 4-month longitudinal deployment with 16 participants, we showed that our system revealed location-dependent differences in communication patterns between home and office environments. Moreover, mixed linear model results further showed significant associations between office-based communication and stress and productivity in this cohort.


\section{Study Methods}

\subsection{Overview}

All study procedures were approved and conducted according to the Institutional Review Board of the University of Southern California, with all participants providing informed consent, and all authors completed the required human subjects research training. 
Participants were recruited from a higher education institution through a Qualtrics screening survey to assess eligibility. Inclusion criteria required that participants work at least 30 hours per week at a computer workstation with a dedicated office workspace and maintain full-time employment for the duration of the study.
Prior to enrollment, participants received a screening interview in which study procedures were explained in detail, and written informed consent was obtained from all participants.
During onboarding session, participants were briefed on all data collection procedures. Each participant was instructed to complete four surveys per workday on their personal smartphone and to wear the wearable audio sensor throughout their working hours. 
Participants were given monetary compensation determined upon \textit{monthly} compliance with each part of the data collection. The data collection lasted for four months for each participant.


\subsection{Study Participants}

\begin{figure*}[ht] {
    \centering
    
    \includegraphics[width=0.86\linewidth]
    {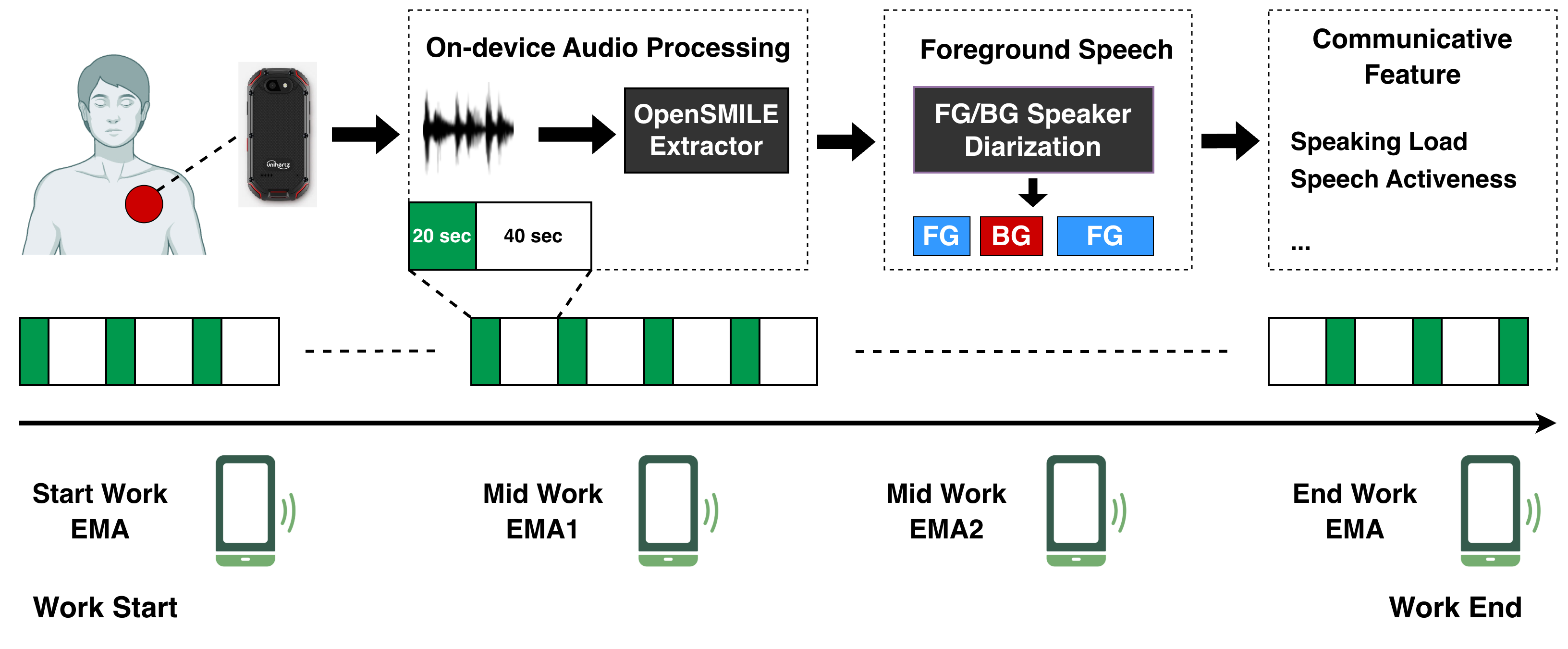}
    \vspace{-4.5mm}

    \caption{
    Overview of the study protocol. EMAs were distributed via text message four times daily, with the first and last surveys aligned to each participant's self-reported work start and end times. Concurrently, participants wore a wearable audio sensor that performs on-device decoding of speech signals into acoustic features, followed by offline foreground speaker extraction and communicative feature computation.}
    \label{fig:wearables}
    \vspace{-3mm}
} \end{figure*}

Table~\ref{tab:demographics} summarizes the demographics and work schedule characteristics of the study participants in this analysis. A total of 19 participants were recruited for the current cohort, of whom 16 wore the audio sensors for more than 10 workdays and were included in this analysis. 
The cohort comprised 10 female and 6 male participants, with a mean age of 43.5 $\pm$ 7.9 years (range: 27–56). 
Participants reported work start times ranging from 06:00 to 09:00 AM, with the majority beginning work around 09:00 AM. On average, participants spent approximately 70.0\% of their working time in the office, and the average work duration was 8.6 hours. Participants were instructed to wear an egocentric audio badge sensor throughout their working hours on each day of the study. Specifically, the detailed design of the system was described in Section~\ref{sec:sensor}. The sensor was built upon the Atom~\cite{atom} device, a compact and cost-effective Android device (Fig.~\ref{fig:wearables}), and ran a custom audio sensing application that performed real-time, on-device acoustic feature extraction via the openSMILE toolkit~\cite{eyben2010opensmile}. 

\begin{table}[t]
    \centering
    \caption{Participant demographic and work schedule characteristics. Office time (\%) represented the self-reported proportion of total working time spent in the office.}
    \label{tab:demographics}
    \begin{tabular}{lccc}
    \toprule
    
    \multirow{2}{*}{\textbf{Demographics}} & \textbf{Female} & \textbf{Male} & \textbf{Total} \\
    
    & \textbf{($n=10$)} & \textbf{($n=6$)} & \textbf{($n=16$)} \\
    \midrule
    \textbf{Age (Years)}  & & & \\
    \hspace{1mm} $\mu\pm\sigma$ & $42.6\pm8.7$ & $40.5\pm7.6$ & $43.5\pm7.9$ \\
    \hspace{1mm} Range & 27-56 & 30-51 & 27-56 \\
    \textbf{Self-report Work Time}  & & & \\
    \hspace{1mm} Start Time (Range) & 8-9AM & 6-9AM & 6-9AM \\
    \hspace{1mm} Duration (hrs) ($\mu\pm\sigma$) & $8.4\pm1.9$ & $9.1\pm3.8$ & $8.6\pm2.7$ \\
    \textbf{Self-report Office Time} & & \\
    \hspace{1mm} \% to Total Time ($\mu\pm\sigma$) & $76.0\pm 20.1$ & $59.2\pm26.8$ & $70.0\pm 22.3$ \\
    \bottomrule
    \end{tabular}
    \vspace{-2mm}
\end{table}

\subsection{Ecological Momentary Assessment}

EMA surveys were distributed via text messages using the Qualtrics application four times per workday (Monday through Friday). The first and last surveys were timed to align with each participant's self-reported work start and end times. The two midday surveys were delivered at randomly generated times within this window. The EMA distribution timeline was illustrated in Figure~\ref{fig:wearables}. Within each EMA, participants reported their current work location from three options: in-office, at home, or other. The present analysis focused specifically on responses corresponding to the two primary occupational contexts, which were the office and the home. Moreover, participants provided subjective assessments of two key psychological outcomes using continuous scales ranging from 0 to 100: perceived productivity (from unproductive to productive) and stress (from no stress to complete stress). 
Overall, these measures offered rich insights into how subjective work experiences varied between the two workspaces.

\section{Wearable Audio Sensing and Computing}
\label{sec:sensor}
The complete system was presented in Figure~\ref{fig:wearables} with two core components: a wearable audio sensor for real-time acoustic feature extraction and an offline speaker detection module to identify the speaker of interest. The outputs were then used to extract speaking load and speaking activeness.

\subsection{Wearable Audio Sensor}

Figure~\ref{fig:wearables} demonstrates the wearable sensor used in this study, which is built on a lightweight and cost-effective Android device called the Atom smartphone. 
Moreover, we develop a customized Android application running on this device that integrates the openSMILE toolkit to extract low-level acoustic features from raw audio. These features, such as F0 (fundamental frequency), loudness, and a set of filterbank coefficients (MFCC 1–12), are well-established features that characterize speech patterns and speaking states. The application processes the acoustic features on-device over a 20-second window at one-minute intervals. We want to highlight that the complete wearable system remains highly cost-effective, with the per-unit cost not exceeding $\$200$ USD.

\subsection{Foreground Speaker Diarization}

\noindent \textbf{Problem Formulation} Since our primary goal is to characterize the communication activity of device-wearing participants, it is essential for us to identify acoustic features that originate primarily from the speaker of interest, namely the participants themselves. To extract audio features of device-wearing participants, for each audio frame, we define \textit{foreground} (FG) speech activity as the speech produced by the device-wearing participants, and \textit{background} (BG) speakers as all other speakers present in the interaction. The other remaining audio is the silence or environmental noise (S). This definition follows the previous work presented in \cite{nadarajan2019speaker} and \cite{liang2023dataset}.

\vspace{1mm}
\noindent \textbf{Model Development}
We build the ResNet-based FG/BG speaker diarization from our previous work~\cite{feng2026voxcare} to predict frame-wise FG/BG speaker labels. Specifically, this model takes filterbank features (MFCC 1-12) as the input to perform FG/BG speaker diarization. Given that collecting raw audio waveforms in the working environment to curate an annotated training corpus for this study is infeasible due to privacy constraints. As a result, we train the proposed FG/BG speaker diarization model using publicly available corpora that contain egocentric speech collected in a similar form factor. In particular, we use three publicly available datasets in our experiments: MMCSG~\cite{zmolikova2024chime}, the ICSI-Meeting Corpus~\cite{janin2003icsi}, and the EasyCom Corpus~\cite{donley2021easycom}. Moreover, as environmental noise is a primary source of degradation in wearable audio systems deployed in the wild, we augment the training data by mixing indoor noise samples from the AudioSet dataset~\cite{gemmeke2017audio}.

\vspace{1mm}
\noindent \textbf{Teacher-student Distillation} Given the data privacy and limitations on on-device computation, only a small set of acoustic features is extracted by the wearable audio sensor for FG speech region inference, which inherently limits model performance. To address this, we introduce a second-stage teacher-student distillation framework.
Specifically, we develop a suite of FG/BG speaker diarization models based on speech foundation models as teachers. The experimented foundation models include the WavLM~\cite{chen2022wavlm} and Whisper model families~\cite{radford2023robust}, each fine-tuned using low-rank adaptation (LoRA~\cite{hu2022lora}). All models share a unified architecture comprising a weighted average layer that aggregates representations across all hidden layers, followed by a 1D convolutional neural network that maps the weighted output to frame-level FG/BG speaker predictions.

\begin{figure} {
    \centering
    
    \includegraphics[width=0.9\linewidth]{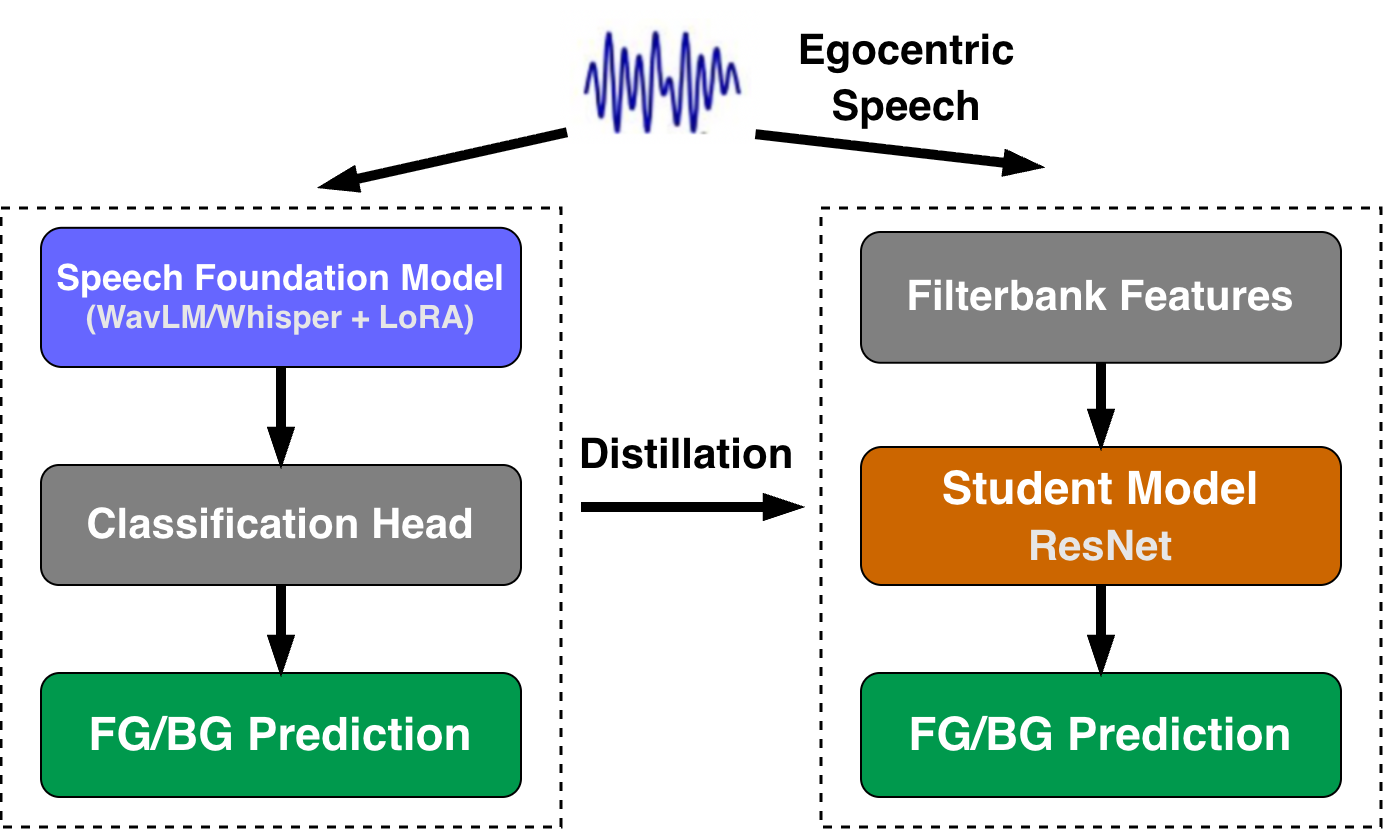}
    \vspace{-0mm}
    \caption{Overview of the FG/BG speaker diarization modeling that relies on teacher–student learning for improved FG/BG speaker diarization. Specifically, a teacher network based on a speech foundation model and a student model is first trained to perform FG/BG diarization. The teacher network is then transferred to the student network through knowledge distillation.}
    \label{fig:ego_fg}
    \vspace{-3mm}
} \end{figure}

In the initialization stage of the distillation, we selected the best-performing model as the teacher model based on the empirical performance. In the distillation process, the weights of the teacher model were frozen while we continued fine-tuning the student model. The training objective combined a frame-wise cross-entropy loss $\mathcal{L}_{CE}$ for classification with a KL-divergence loss $\mathcal{L}_{KLD}$ that encourages the student's outputs to align with those of the teacher: $\mathcal{L}_{total} = \mathcal{L}_{CE}(p_{s}) + \alpha \mathcal{L}_{KLD}(p_{t}, p_{s})$, where $s$ and $t$ stand for student and teacher, respectively. We aim to use this distillation process to further improve the FG/BG speaker diarization model that operates directly on low-dimensional acoustic features.

\subsection{Communication Behavior Extraction}

Here, we described the computation of communicative features from the retained FG speech regions. 
A minute was considered to contain valid speech activity if at least 250 FG speech frames were detected within the corresponding 20-second feature recording window.

\vspace{0.5mm}
\noindent \textbf{Speaking Load} Speaking time per hour was computed as a measure of communication load during work. We deliberately excluded total speaking time, as individuals differed in their overall working hours, which would otherwise introduce systematic confounds related to work schedule rather than average communicative behavior.

\vspace{0.5mm}
\noindent \textbf{Speaking Activeness}
Apart from communication load, we estimated activeness, or arousal, from speech using a rule-based framework~\cite{bone2014robust} to characterize how actively individuals engage in communication throughout the work. 
This method was particularly well-suited to applications where collecting labeled training data was not feasible. Activeness was estimated from two acoustic features: log-pitch and intensity. The framework first constructed a personalized empirical distribution $N_{i}$ for each feature type $i\in\{1, 2\}$ to account for inter-speaker variability. Then, the activeness score $p_{i, j}$ for feature type $i$ in the $j$-th recording was then derived as the median value of that feature relative to its personalized empirical distribution $N_{i}$, given by $p_{i, j} = 2 \times E[x_{i, j} > N_{i}] - 1$.

To improve estimation robustness, the individual feature scores were combined via a weighted fusion, where each weight is the Spearman rank correlation $r_{i}$ between the score vector $p_{i}$ and the mean score vector $p_{\mu}$ across all recordings. We defined the correlation vector as $\mathbf{r} = (r_{1}, r_{2})$, and the weights were then normalized to have unit combined magnitude: $w_{i} = \frac{r_{i}}{||\mathbf{r}||}$. As~\cite{bone2014robust} suggested, positive scores indicated relatively more active states compared to each participant's personal baseline. We summarized speaking activeness within each working day using the 90th percentile of all scores. This feature captured the peak communicative activation of the participant at work.

\section{FG/BG Diarization Experiments}
\label{sec:fg_experiments}

\subsection{Datasets} 

Three egocentric audio datasets collected with comparable form factors were used in this work. 
The first two datasets were recorded using the smart glasses: the MMCSG corpus~\cite{zmolikova2024chime} and the EasyCom corpus~\cite{donley2021easycom}. Particularly, MMCSG consisted of 530 unique three-minute dyadic interaction recordings from 138 participants, collected through Aria smart glasses worn by one subject. EasyCom included 12 sessions totaling approximately five hours, each recording small groups of three to five participants engaged in office-like conversations of roughly 30 minutes. In contrast, the ICSI meeting corpus~\cite{janin2003icsi} captured naturalistic multi-party conversations using environmentally placed microphones that support both close-talk and far-field conditions. 
For all datasets, manual speaker boundary annotations were available to derive ground-truth FG, BG, and silence regions. Model training was performed over non-overlapping 10-second windows.

\begin{table*}[t]
    \centering

    \caption{Comparison of FG/BG speaker diarization results in our proposed work. VAD, SC, and DER represent voice activity detection error rate, speaker confusion rate, and diarization error rate, respectively. The lower number represents better performance.}
    \vspace{-1mm}
    \resizebox{\linewidth}{!}{
        \begin{tabular}{lccccccccc}
    
            \toprule
    
            \multirow{2}{*}{\textbf{Model}} & 
            \multicolumn{3}{c}{\textbf{MMCSG}} & 
            \multicolumn{3}{c}{\textbf{ICSI-Meeting Corpus}} & 
            \multicolumn{3}{c}{\textbf{EasyCom}}
            \\
            
            & \multicolumn{1}{c}{\textbf{VAD}$\downarrow$} & \multirow{1}{*}{\textbf{SC}$\downarrow$} & 
            \multicolumn{1}{c}{\textbf{DER}$\downarrow$} & \multirow{1}{*}{\textbf{VAD}$\downarrow$} &
            \multicolumn{1}{c}{\textbf{SC}$\downarrow$} & \multirow{1}{*}{\textbf{DER}$\downarrow$} & 
            \multicolumn{1}{c}{\textbf{VAD}$\downarrow$} & \multirow{1}{*}{\textbf{SC}$\downarrow$} & 
            \multirow{1}{*}{\textbf{DER}$\downarrow$} \\
             
            \cmidrule(lr){1-1}
            \cmidrule(lr){2-4}
            \cmidrule(lr){5-7}
            \cmidrule(lr){8-10}
    
            \rowcolor{gray!10}
            \textbf{Teacher Model} \\
    
            \hspace{1mm} WavLM-Large &
            7.7$_{\pm 0.2}$ & 
            \textbf{0.7}$_{\pm 0.1}$ & 
            8.4$_{\pm 0.2}$ & 
            13.0$_{\pm 0.3}$ &
            0.3$_{\pm 0.1}$ & 
            13.3$_{\pm 0.4}$ & 
            24.7$_{\pm 1.4}$ & 
            \textbf{0.7}$_{\pm 0.3}$ & 
            25.4$_{\pm 0.5}$
            \\
            
            \hspace{1mm} Whisper-Tiny &
            14.5$_{\pm 0.5}$ & 
            4.0$_{\pm 0.1}$ & 
            18.5$_{\pm 0.5}$ & 
            17.5$_{\pm 0.8}$ &
            1.9$_{\pm 0.3}$ & 
            19.4$_{\pm 0.5}$ & 
            31.7$_{\pm 1.5}$ & 
            2.2$_{\pm 0.4}$ & 
            33.9$_{\pm 3.0}$
            \\
    
            \hspace{1mm} Whisper-Small &
            7.1$_{\pm 0.1}$ & 
            0.8$_{\pm 0.1}$ &
            7.9$_{\pm 0.1}$ & 
            10.0$_{\pm 0.2}$ &
            0.2$_{\pm 0.1}$ & 
            10.2$_{\pm 0.2}$ & 
            23.1$_{\pm 0.4}$ & 
            1.0$_{\pm 0.1}$ & 
            24.1$_{\pm 0.5}$ \\
    
            \hspace{1mm} Whisper-Large &
            \textbf{6.5}$_{\pm 0.2}$ & 
            0.8$_{\pm 0.1}$ & 
            \textbf{7.3}$_{\pm 0.3}$ & 
            \textbf{8.3}$_{\pm 0.2}$ &
            \textbf{0.1}$_{\pm 0.1}$ & 
            \textbf{8.4}$_{\pm 0.3}$ & 
            \textbf{21.6}$_{\pm 0.6}$ & 
            1.0$_{\pm 0.1}$ & 
            \textbf{22.6}$_{\pm 0.5}$ \\

            \rowcolor{gray!10}
            \textbf{Student Model} \\
    
            \hspace{1mm} ResNet &
            11.2$_{\pm 0.2}$ & 
            4.8$_{\pm 0.1}$ & 
            16.0$_{\pm 0.4}$ & 
            23.4$_{\pm 1.0}$ &
            1.2$_{\pm 0.2}$ & 
            24.6$_{\pm 0.9}$ & 
            28.5$_{\pm 1.5}$ & 
            3.3$_{\pm 0.2}$ & 
            31.8$_{\pm 1.3}$ \\

            \hspace{1mm} Distilled-ResNet &
            \textbf{10.3}$_{\pm 0.3}$ & 
            \textbf{3.8}$_{\pm 0.7}$ & 
            \textbf{14.1}$_{\pm 0.8}$ &
            \textbf{22.5}$_{\pm 0.6}$ &
            \textbf{1.0}$_{\pm 0.2}$ & 
            \textbf{23.5}$_{\pm 0.7}$ & 
            \textbf{26.7}$_{\pm 0.9}$ & 
            \textbf{2.3}$_{\pm 0.3}$ & 
            \textbf{29.0}$_{\pm 0.3}$ \\
    
            \bottomrule
    
        \end{tabular}
    }
    \label{tab:fg_pred}
    \vspace{-3.5mm}
\end{table*}

\subsection{Experimental Details}

All FG/BG diarization models were trained on a high-performance computing (HPC) cluster using a collective training set combining three datasets. 
For MMCSG, we adopted the default train/validation/test split, while for the remaining datasets, recording sessions were partitioned into 60\%, 20\%, and 20\% for training, validation, and testing, respectively.
All models were optimized with a learning rate of $5\times10^{-5}$ over 10 epochs. The knowledge distillation weight $\alpha$ was empirically selected from \{1, 2, 10\}, $\alpha=2$ yielding the best results. During training, noise augmentation was applied by mixing office-related noise with clean audio at signal-to-noise ratios from 0 to 30 dB. During inference, the model predicted the frame-wise speaker labels every 10 seconds, which were then merged into speaker segments for computing the diarization metric.
We reported the overall diarization error rate (DER) along with the speaker confusion rate (SC) and voice activity detection error rate (VAD).

\begin{figure} {
    \centering
    
    \includegraphics[width=\linewidth]{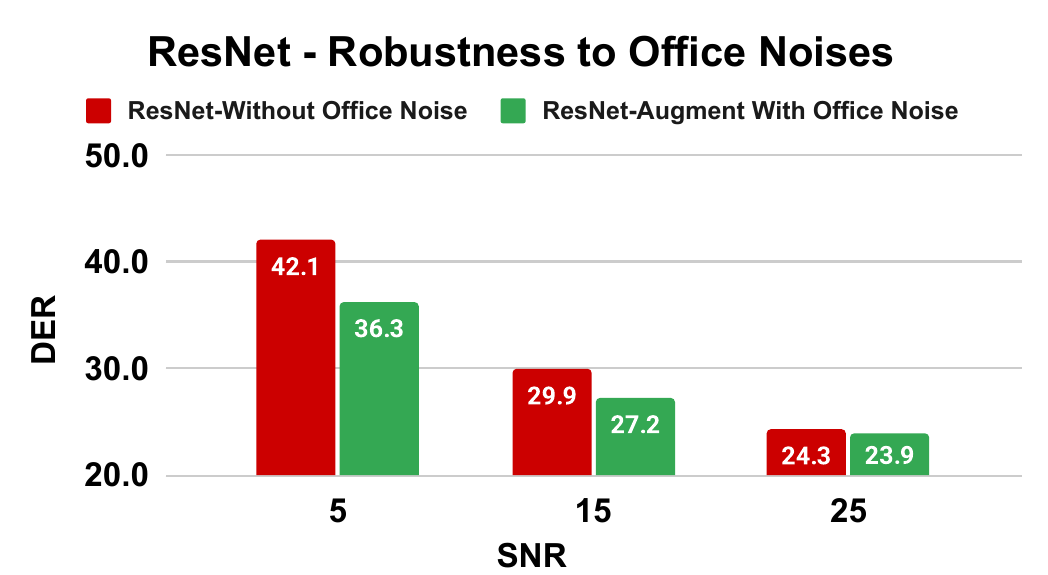}
    \vspace{-6.5mm}
    \caption{Evaluation of Distilled-ResNet models trained with and without office noise augmentation. We reported average DER across three datasets at Signal-to-noise ratios of 5, 15, and 25 dB.}
    \label{fig:fg_pred_noise}
    \vspace{-3mm}
} \end{figure}

\subsection{FG/BG Diarization Results}

\noindent \textbf{Benchmark Performance} Table~\ref{tab:fg_pred} presents the evaluation of FG/BG speaker diarization across three benchmark datasets under both teacher and student model setups. Among the teacher models, Whisper-Large consistently achieved the strongest overall performance, while Whisper-Small and WavLM-Large demonstrate competitive results, particularly in minimizing speaker confusion on MMCSG and EasyCom. Meanwhile, Whisper-Tiny, the smallest teacher model in the comparison, substantially underperformed all other teacher models. In the student model, the proposed Distilled-ResNet consistently outperformed the baseline ResNet across all metrics and datasets, achieving DER reductions of 1.9, 1.1, and 2.8 points on MMCSG, ICSI, and EasyCom, respectively. Particularly, Distilled-ResNet achieved comparable performance to Whisper-Tiny, which was pre-trained on millions of hours of speech data. These results highlighted that knowledge distillation from large-scale speech foundation models into compact architectures was an effective and practical strategy for scenarios where raw audio was inaccessible.

\vspace{0.5mm}
\noindent \textbf{Model Performance against Environmental Noises} Since egocentric audio recordings in naturalistic settings were inherently coupled with environmental noise, it was critical to evaluate the robustness of our model against the model trained without indoor noise augmentation. Specifically, for the test set of each dataset, we mixed it with indoor noise samples from the AudioSet~\cite{gemmeke2017audio} dataset at varying SNR levels. Figure~\ref{fig:fg_pred_noise} reports the average DER across all three datasets at three SNR levels. Noise augmentation during training yielded consistent improvements under low-SNR conditions, with DER reductions of 5.8 and 2.7 percent at SNR 5 and 15, respectively. At higher SNR=25, the two models achieved comparable performance. This suggested that noise augmentation provided the greatest benefit in the low-SNR condition, which reflected the challenges in real-world naturalistic recordings.

\begin{figure} {
    \centering
    
    \includegraphics[width=\linewidth]{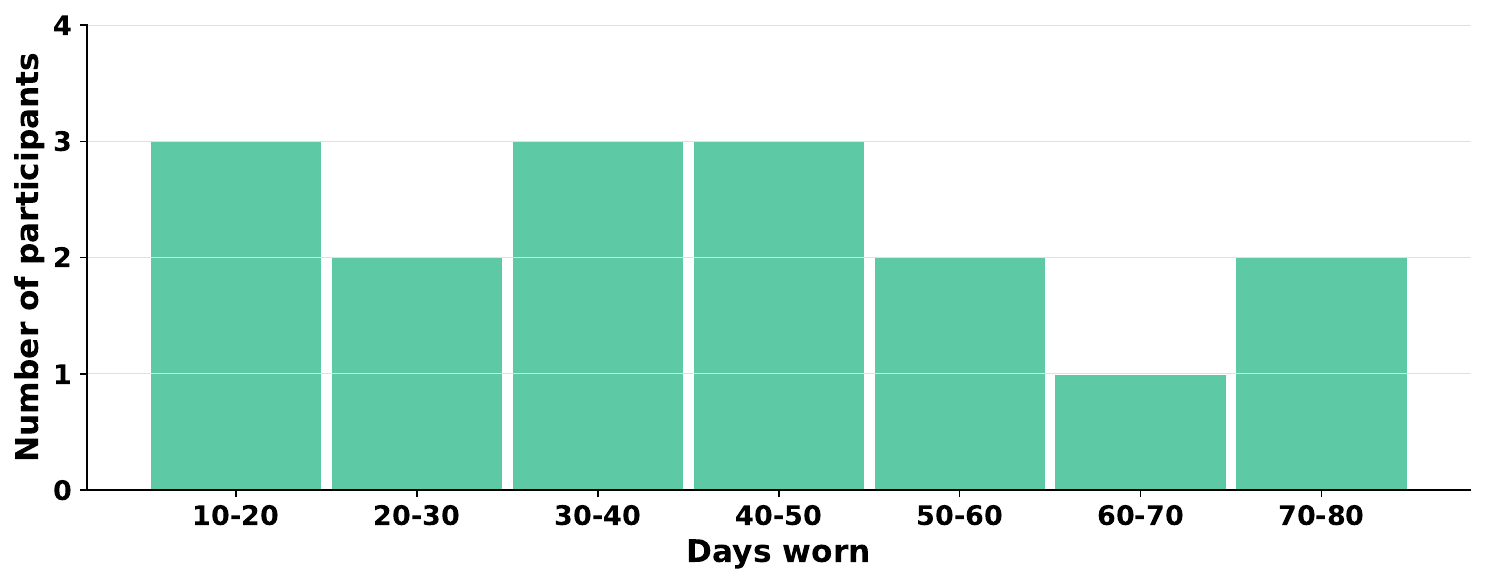}
    \vspace{-7mm}
    \caption{Overview of sensor wearing statistics.}
    \label{fig:sensor}
    \vspace{-3mm}
} \end{figure}

\section{Communication Patterns}

In this section, we analyzed communication patterns of participants between working in the office and at home. This analysis was based on the FG speech regions inferred by the Distilled-ResNet FG/BG diarization model shown in Table~\ref{tab:fg_pred}. Increasingly, we applied a filtering criterion to keep only frames with a fundamental frequency greater than zero, which further excluded likely unvoiced segments to prevent false predictions given the larger VAD errors shown in Table~\ref{tab:fg_pred}.
The sensor wearing distribution is illustrated in Figure~\ref{fig:sensor}. Within this cohort, participants wore the audio sensor for an average of 39.6$\pm$16.1 days at work, with individual wearing durations ranging from 13 to 74 days. The distribution of wearing days was broadly spread, with the majority of participants wearing the sensor between 30 and 50 days.

\subsection{Communication Patterns between Home and Office}

\begin{figure} {
    \centering
    
    \includegraphics[width=\linewidth]{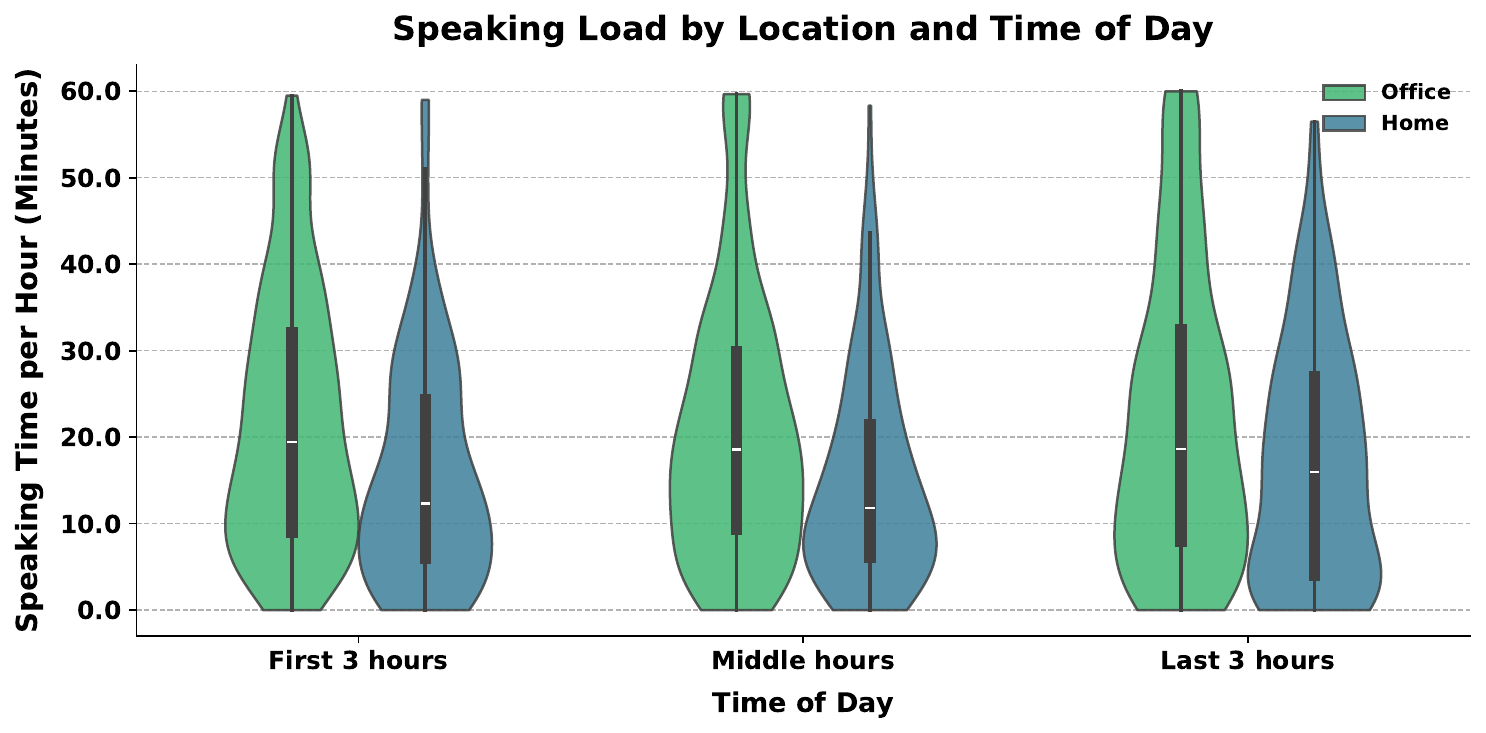}
    \vspace{-7mm}
    \caption{Distribution of speaking time per hour across three workday periods by work location (home vs. office). Participants showed higher speaking time per hour when working in the office across all work periods.}
    \label{fig:speaking_location_time}
    \vspace{-3mm}
} \end{figure}

\vspace{1mm}
\noindent \textbf{Speaking Load} We applied a mixed-effects linear model to examine differences in speaking time per hour between home and office environments, with participants included as a random effect. We added age and sex as the fixed-effect covariates.
From the results shown in Table~\ref{tab:lmm_speaking_arousal}, location was the only significant predictor of speaking time per hour, with participants speaking significantly more at the office compared to home ($\beta$=12.61, 95\% CI [10.7, 14.5], p$<$0.01). Estimated marginal means at the sample's mean age revealed that the speaking time per hour in the office is twice as much as working at home. Meanwhile, neither sex nor age shows statistical significance in predicting communication load. We further plotted the speaking time per hour in the first 3 hours, middle hours, and last 3 hours at work in Fig.~\ref{fig:speaking_location_time}. Across all three work periods, office-based work consistently showed higher speaking rates compared to working at home. This suggested the richer communicative opportunities in the office environments throughout a workday. Particularly, speaking minutes appeared to increase slightly toward the latter part of the workday in both locations. This pattern may reflect end-of-day wrap-up behaviors, such as status updates and coordination of pending tasks, which naturally drive communicative activity upward as the workday draws to an end.

\begin{table}[t]
    \centering
    \caption{Mixed linear model regression results for speaking load and activeness as dependent variables. Bold p-values indicated statistical significance ($p<0.05$).}
    \vspace{-2mm}
    \begin{tabular}{lcccc}
        \toprule
        & \multicolumn{2}{c}{\textbf{Speaking Load}} & \multicolumn{2}{c}{\textbf{Speaking Activeness}} \\
        \textbf{Predictor} & \textbf{Coef.} & \textbf{p-value} & \textbf{Coef.} & \textbf{p-value} \\
        \cmidrule(lr){1-1} \cmidrule(lr){2-3} \cmidrule(lr){4-5}
        Intercept & 1.59   & 0.912 & -0.17  & \textbf{0.026} \\
        Sex (Male) & 4.54   & 0.384             & 0.01   & 0.750 \\
        Location (Office)   & 12.61  & \textbf{$<$0.001} & 0.12   & \textbf{$<$0.001} \\
        Age & 0.15 & 0.626 & 0.00   & 0.083 \\
        \cmidrule(lr){1-1} \cmidrule(lr){2-3} \cmidrule(lr){4-5}
        Group Var  & 103.13 & - & 0.01   & - \\
        \bottomrule
    \end{tabular}
    \vspace{-3mm}
    \label{tab:lmm_speaking_arousal}
\end{table}

\vspace{0.5mm}
\noindent \textbf{Speaking Activeness} We applied a similar mixed-effects linear model to explore the differences in communication activeness between working at home and in the office. The results in Table~\ref{tab:lmm_speaking_arousal} revealed that office location is a significant positive predictor of speaking activeness per workday ($\beta$= 0.12, p$<$0.001), indicating that participants were more active in communication when working in the office compared to at home. In contrast, neither sex nor age was statistically significant in the modeling.

\begin{table}[t]
    \centering
    \caption{Mixed linear model results examining the effects of location-specific slope term for communication patterns on stress and productivity. Bold indicates statistical significance ($p<0.05$).}
    \vspace{-1mm}
    \begin{tabular}{lcccc}
        \toprule
        & \multicolumn{2}{c}{\textbf{Stress}} & \multicolumn{2}{c}{\textbf{Productivity}} \\

        & \multicolumn{2}{c}{\textbf{(n=1668)}} & \multicolumn{2}{c}{\textbf{(n=1793)}} \\

        \textbf{Predictor} & \textbf{Coef.} & \textbf{p-value} & \textbf{Coef.} & \textbf{p-value} \\
        \cmidrule(lr){1-1} \cmidrule(lr){2-3} \cmidrule(lr){4-5}
        Intercept & 31.54 & \textbf{0.001} & 49.87 & \textbf{$<$0.001} \\
        Sex (Male) & 12.48  & 0.079 & 4.42 & 0.369 \\
        Survey (Mid-work EMA) & -2.10 & 0.263 & 0.04   & 0.983 \\
        Survey (End-work EMA) & -1.06 & 0.599 & 0.23   & 0.903 \\
        Location (Office) & 2.93 & 0.395 & 18.24  & \textbf{$<$0.001} \\
        Age & -8.74  & 0.485 & 7.73 & 0.371 \\
        Speaking Load $\times$ Home   & 3.13 & 0.388 & 1.40 & 0.681 \\
        Speaking Load $\times$ Office  & 7.49 & \textbf{0.004} & -5.87  & \textbf{0.016} \\
        Speaking Activeness $\times$ Home & 9.23   & \textbf{0.030} & 14.07  & \textbf{$<$0.001} \\
        Speaking Activeness $\times$ Office & 4.60 & 0.239 & -0.34  & 0.926 \\
        \cmidrule(lr){1-1} \cmidrule(lr){2-3} \cmidrule(lr){4-5}
        Group Var & 191.36 & - & 90.03  & - \\
        \bottomrule
    \end{tabular}
    \vspace{-3mm}
    \label{tab:lmm_combined}
\end{table}

\subsection{Communication Patterns and Self-report EMAs}

We applied a mixed linear model to jointly examine the effects of speaking activeness and load on self-reported stress and productivity. 
Specifically, we extracted these two features within a 2-hour window prior to each EMA completion. All features were z-normalized before processing in the mixed linear model.
We also added location-specific slope terms for speaking load and activeness, with participants as random effects. Location (home vs. office), survey time point (start-work, mid-work, and end-work), age, and sex were included as covariates. As shown in Table~\ref{tab:lmm_combined}, office location was a strong positive predictor of productivity ($\beta$ = 18.24, p$<$0.001), suggesting that the office environment was inherently associated with higher self-report productivity scores. With respect to speaking load, greater speech per hour in the office was associated with higher self-report stress ($\beta$ = 7.49, p=0.004) and lower self-report productivity ($\beta$ = -5.87, p = 0.016), implying that excessive communicative demands in the office context may lead to higher affective loads. Regarding speaking activeness, its interaction with home location was a significant positive predictor of both stress ($\beta$ = 9.23, p=0.030) and productivity ($\beta$ = 14.07, p$<$0.001), hinting that increased speaking activeness during home-based work may amplify both productivity and affective load. In contrast, neither the office-specific slope for speaking activeness and the home-specific slope for speaking load were statistically significant for either outcome. Overall, these findings highlighted the location-specific interaction between communication patterns and self-report affective outcomes.

\section{Conclusion}

In this work, we presented an egocentric wearable audio sensing system that derived interpretable communicative measures, including speaking load and activeness, through on-device acoustic feature extraction and offline FG/BG speaker diarization. To address the inherent limitations of low-dimensional acoustic features available on-device, we used a teacher-student knowledge distillation framework that transferred knowledge from large-scale speech foundation models into a compact ResNet model, achieving consistent improvements over the baseline ResNet and reaching comparable performance to Whisper-Tiny. 
In a four-month longitudinal deployment among office workers, our system revealed systematic location-dependent differences in communication patterns between home and office environments. The mixed linear model results further demonstrated significant associations between office-based communication and self-reported stress and productivity. Future work will expand the participant cohort and integrate multimodal signals to enrich the characterization of hybrid work experiences.

\section*{AI-Generated Content Disclosure}
Generative AI assisted with the formatting and grammar correction of this paper. All experimental design, data collection, and analysis were performed by the human authors.

\bibliographystyle{IEEEtranS}
\bibliography{refs}

@article{fukumura2021negotiating,
  title={Negotiating time and space when working from home: experiences during COVID-19},
  author={Fukumura, Yoko E and Schott, Joseph M and Lucas, Gale M and Becerik-Gerber, Burcin and Roll, Shawn C},
  journal={OTJR: Occupation, Participation and Health},
  volume={41},
  number={4},
  pages={223--231},
  year={2021},
  publisher={Sage Publications Sage CA: Los Angeles, CA}
}

@article{awada2021working,
  title={Working from home during the COVID-19 pandemic: Impact on office worker productivity and work experience},
  author={Awada, Mohamad and Lucas, Gale and Becerik-Gerber, Burcin and Roll, Shawn},
  journal={Work},
  volume={69},
  number={4},
  pages={1171--1189},
  year={2021},
  publisher={SAGE Publications Sage UK: London, England}
}

@article{yang2022effects,
  title={The effects of remote work on collaboration among information workers},
  author={Yang, Longqi and Holtz, David and Jaffe, Sonia and Suri, Siddharth and Sinha, Shilpi and Weston, Jeffrey and Joyce, Connor and Shah, Neha and Sherman, Kevin and Hecht, Brent and others},
  journal={Nature human behaviour},
  volume={6},
  number={1},
  pages={43--54},
  year={2022},
  publisher={Nature Publishing Group UK London}
}

@article{donley2021easycom,
  title={Easycom: An augmented reality dataset to support algorithms for easy communication in noisy environments},
  author={Donley, Jacob and Tourbabin, Vladimir and Lee, Jung-Suk and Broyles, Mark and Jiang, Hao and Shen, Jie and Pantic, Maja and Ithapu, Vamsi Krishna and Mehra, Ravish},
  journal={arXiv preprint arXiv:2107.04174},
  year={2021}
}

@inproceedings{janin2003icsi,
  title={The ICSI meeting corpus},
  author={Janin, Adam and Baron, Don and Edwards, Jane and Ellis, Dan and Gelbart, David and Morgan, Nelson and Peskin, Barbara and Pfau, Thilo and Shriberg, Elizabeth and Stolcke, Andreas and others},
  booktitle={2003 IEEE International Conference on Acoustics, Speech, and Signal Processing, 2003. Proceedings.(ICASSP'03).},
  volume={1},
  pages={I--I},
  year={2003},
  organization={IEEE}
}

@inproceedings{zmolikova2024chime,
  title={The CHiME-8 MMCSG Challenge: Multi-modal conversations in smart glasses},
  author={Zmolikova, Katerina and Merello, Simone and Kalgaonkar, Kaustubh and Lin, Ju and Moritz, Niko and Ma, Pingchuan and Sun, Ming and Chen, Honglie and Saliou, Antoine and Petridis, Stavros and others},
  booktitle={8th International Workshop on Speech Processing in Everyday Environments (CHiME)},
  pages={7--12},
  year={2024}
}

@article{hu2022lora,
  title={Lora: Low-rank adaptation of large language models.},
  author={Hu, Edward J and Shen, Yelong and Wallis, Phillip and Allen-Zhu, Zeyuan and Li, Yuanzhi and Wang, Shean and Wang, Lu and Chen, Weizhu and others},
  journal={ICLR},
  volume={1},
  number={2},
  pages={3},
  year={2022}
}

@inproceedings{eyben2010opensmile,
  title={Opensmile: the munich versatile and fast open-source audio feature extractor},
  author={Eyben, Florian and W{\"o}llmer, Martin and Schuller, Bj{\"o}rn},
  booktitle={Proceedings of the 18th ACM international conference on Multimedia},
  pages={1459--1462},
  year={2010}
}

@INPROCEEDINGS{Wearable, 
    author={T. Martin and E. Jovanov and D. Raskovic}, 
    booktitle={Digest of Papers. Fourth International Symposium on Wearable Computers}, 
    title={Issues in wearable computing for medical monitoring applications: a case study of a wearable ECG monitoring device}, 
    year={2000}, 
    pages={43-49}, 
    month={October}
}

@misc{atom,
    title = {Atom},
    url = {https://www.unihertz.com/products/atom},
}

@article{chen2022wavlm,
  title={Wavlm: Large-scale self-supervised pre-training for full stack speech processing},
  author={Chen, Sanyuan and Wang, Chengyi and Chen, Zhengyang and Wu, Yu and Liu, Shujie and Chen, Zhuo and Li, Jinyu and Kanda, Naoyuki and Yoshioka, Takuya and Xiao, Xiong and others},
  journal={IEEE Journal of Selected Topics in Signal Processing},
  volume={16},
  number={6},
  pages={1505--1518},
  year={2022},
  publisher={IEEE}
}

@inproceedings{radford2023robust,
  title={Robust speech recognition via large-scale weak supervision},
  author={Radford, Alec and Kim, Jong Wook and Xu, Tao and Brockman, Greg and McLeavey, Christine and Sutskever, Ilya},
  booktitle={International conference on machine learning},
  pages={28492--28518},
  year={2023},
  organization={PMLR}
}

@inproceedings{feng2018tiles,
    title={Tiles audio recorder: An unobtrusive wearable solution to track audio activity},
    author={T. Feng and A. Nadarajan and C. Vaz and B. Booth and S. Narayanan},
    booktitle={Proceedings of the 4th ACM Workshop on Wearable Systems and Applications},
    pages={33--38},
    year={2018},
    organization={ACM}
}

@article{bone2014robust,
    title={Robust unsupervised arousal rating: A rule-based framework withknowledge-inspired vocal features},
    author={D. Bone and C. Lee and S. Narayanan},
    journal={IEEE transactions on affective computing},
    volume={5},
    number={2},
    pages={201--213},
    year={2014},
    publisher={IEEE}
}

@inproceedings{nadarajan2019speaker,
    author = {A. Nadarajan and K. Somandepalli and S. Narayanan},
    booktitle = {Proceedings of ICASSP},
    location = {Brighton, UK},
    month = {May},
    title = {Speaker Agnostic Foreground Speech Detection From Audio Recordings in Workplace Settings From Wearable Recorders},
    year = {2019}
}

@article{feng2026voxcare,
  title={VoxCare: Studying Natural Communication Behaviors of Hospital Caregivers through Wearable Sensing of Egocentric Audio},
  author={Feng, Tiantian and Avramidis, Kleanthis and Xu, Anfeng and Wang, Deqi and Booth, Brandon M and Narayanan, Shrikanth},
  journal={arXiv preprint arXiv:2603.10888},
  year={2026}
}

@article{mundnich2020tiles2018,
  title={TILES-2018: A longitudinal physiologic and behavioral data set of hospital workers},
  author={Karel Mundnich and Brandon M. Booth and Michelle L'Hommedieu and Tiantian Feng and Benjamin Girault and Justin L'Hommedieu and Mackenzie Wildman and Sophia Skaaden and Amrutha Nadarajan and Jennifer L. Villatte and Tiago H. Falk and Kristina Lerman and Emilio Ferrara and Shrikanth Narayanan},
  journal={Sci Data},
  volume={7},
  number={354},
  year={2020},
}

@article{walther2005rules,
  title={The rules of virtual groups: Trust, liking, and performance in computer-mediated communication},
  author={Walther, Joseph B and Bunz, Ulla},
  journal={Journal of communication},
  volume={55},
  number={4},
  pages={828--846},
  year={2005},
  publisher={Wiley Online Library}
}

@article{yau2022tiles,
  title={TILES-2019: A longitudinal physiologic and behavioral data set of medical residents in an intensive care unit},
  author={Yau, Joanna C and Girault, Benjamin and Feng, Tiantian and Mundnich, Karel and Nadarajan, Amrutha and Booth, Brandon M and Ferrara, Emilio and Lerman, Kristina and Hsieh, Eric and Narayanan, Shrikanth},
  journal={Scientific Data},
  volume={9},
  number={1},
  pages={536},
  year={2022},
  publisher={Nature Publishing Group UK London}
}

@article{koohsari2021changes,
  title={Changes in workers’ sedentary and physical activity behaviors in response to the COVID-19 pandemic and their relationships with fatigue: longitudinal online study},
  author={Koohsari, Mohammad Javad and Nakaya, Tomoki and McCormack, Gavin R and Shibata, Ai and Ishii, Kaori and Oka, Koichiro},
  journal={JMIR public health and surveillance},
  volume={7},
  number={3},
  pages={e26293},
  year={2021},
  publisher={JMIR Publications Toronto, Canada}
}

@article{fukushima2021associations,
  title={Associations of working from home with occupational physical activity and sedentary behavior under the COVID-19 pandemic},
  author={Fukushima, Noritoshi and Machida, Masaki and Kikuchi, Hiroyuki and Amagasa, Shiho and Hayashi, Toshio and Odagiri, Yuko and Takamiya, Tomoko and Inoue, Shigeru},
  journal={Journal of occupational health},
  volume={63},
  number={1},
  pages={e12212},
  year={2021},
  publisher={Oxford University Press}
}

@inproceedings{gemmeke2017audio,
  title={Audio set: An ontology and human-labeled dataset for audio events},
  author={Gemmeke, Jort F and Ellis, Daniel PW and Freedman, Dylan and Jansen, Aren and Lawrence, Wade and Moore, R Channing and Plakal, Manoj and Ritter, Marvin},
  booktitle={2017 IEEE international conference on acoustics, speech and signal processing (ICASSP)},
  pages={776--780},
  year={2017},
  organization={IEEE}
}

@inproceedings{liang2023dataset,
  title={A dataset for foreground speech analysis with smartwatches in everyday home environments},
  author={Liang, Dawei and Xu, Zifan and Chen, Yinuo and Adaimi, Rebecca and Harwath, David and Thomaz, Edison},
  booktitle={2023 IEEE International Conference on Acoustics, Speech, and Signal Processing Workshops (ICASSPW)},
  pages={1--5},
  year={2023},
  organization={IEEE}
}

@article{viererbl2022you,
  title={“You don’t meet anybody when walking from the living room to the kitchen”: informal communication during remote work},
  author={Viererbl, Benno and Denner, Nora and Koch, Thomas},
  journal={Journal of communication management},
  volume={26},
  number={3},
  pages={331--348},
  year={2022},
  publisher={Emerald Publishing Limited}
}

\end{document}